\documentclass[prb,lengthcheck]{revtex4-1}

\usepackage{amsmath}
\usepackage{bm}
\usepackage{amssymb}
\usepackage{graphicx}
\usepackage[usenames,dvipsnames]{xcolor}
\usepackage{hyperref}
\usepackage{braket}
\usepackage{textcomp}
\usepackage{tikz}
\usepackage{ragged2e}
\usepackage[utf8]{inputenc}
\usepackage{wrapfig}
\usepackage{epstopdf}

\hypersetup{
  colorlinks   = true, 
  urlcolor     = blue, 
  linkcolor    = blue, 
  citecolor   = red, 
}

\newcommand{\tr}{\operatorname{tr}}

\newcommand{\ic}{\ensuremath{\mathrm{i}}}
\newcommand{\rket}[1]{\ket{#1}}
\newcommand{\rbra}[1]{\bra{#1}}
\renewcommand{\vec}[1]{\ensuremath{\bm{#1}}}

\begin{document}

\title{Variational Matrix Product Ansatz for Nonuniform Dynamics in the Thermodynamic Limit}

\author{Ashley Milsted$^{1}$}
\author{Jutho Haegeman$^{2}$}
\author{Tobias J.\ Osborne$^{1}$} 
\author{Frank Verstraete$^{2}$} 
\affiliation{$^1$Leibniz Universit\"at Hannover, Institute of Theoretical Physics, Appelstrasse 2, D-30167 Hannover, Germany\\
$^2$Vienna Center for Quantum Science and Technology, Faculty of Physics, University of Vienna, Boltzmanngasse 5, A-1090 Wien, Austria\\
}


\begin{abstract}
We describe how to implement the time-dependent variational principle for matrix 
product states in the thermodynamic limit for nonuniform lattice systems. This 
is achieved by confining the nonuniformity to a (dynamically expandable) finite 
region with fixed boundary conditions. The suppression of nonphysical 
quasiparticle reflections from the boundary of the nonuniform region is also 
discussed. Using this algorithm we study the dynamics of localized excitations 
in infinite systems, which we illustrate in the case of the spin-1 
anti-ferromagnetic Heisenberg model and the $\phi^4$ model. 
\end{abstract}

\maketitle

Douglas Adams (nearly) put it best: ``[Hilbert] space is big. ... You just won't 
believe how vastly hugely mindbogglingly big it is. I mean, you may think it's 
a long way down the road to the chemist, but that's just peanuts compared to 
[Hilbert] space.'' Given said space's exponential growth with the size of a 
many-particle system, it is a little astounding that general techniques exist 
to allow efficient numerical calculations in a wide range of physically 
interesting cases. This is possible because physically relevant states have 
limited entanglement \cite{eisert_2010, hastings_2007, osborne_2006}. This 
observation may be exploited to obtain an efficient parametrization of this 
\emph{physical corner} of Hilbert space. 

The class of \emph{matrix product states} (MPS) \cite{fannes_finitely_1992, 
*verstraete_matrix_2008, *cirac_renormalization_2009} represents, in one 
dimension, a good parametrization of the physical corner. This is amply 
demonstrated by the unparalleled success of the \emph{density matrix 
renormalization group} (DMRG) \cite{white_density_1992, 
*schollwock_density-matrix_2005}, which can be viewed as a variational method 
when formulated in the MPS language \cite{rommer_class_1996, *schollwock_density-matrix_2011}. 
The MPS class has served as the basis for many exciting generalizations, including 
the study of non-equilibrium dynamics \cite{vidal_2004} and higher-dimensional 
systems \cite{verstraete_cirac_2004}. More recently, Haegeman \emph{et al.}\ 
have implemented the \emph{time-dependent variational principle} (TDVP | see 
boxout) for MPS \cite{haegeman_time-dependent_2011}, providing a locally optimal 
(in time) framework for simulating dynamics, including finding ground states via
imaginary time evolution, and an ansatz for studying excitations of one-dimensional 
lattice systems.

\begin{figure}[b]
  \vspace{-10pt}
  \fbox{
      \begin{minipage}{0.45\textwidth}
        \textbf{The time-dependent variational principle} \\
        \vspace{2pt}
        \justifying
        \begin{wrapfigure}{r}{4.0cm}
          \vspace{-10pt}
          \hspace{-40pt}
          \includegraphics{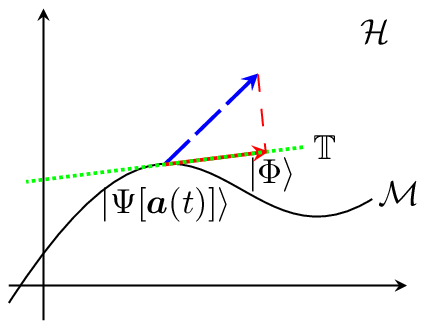}
          \hspace{-35pt}
        \end{wrapfigure}
        A variational manifold $\mathcal{M}$ is depicted as embedded in a 
        Hilbert space $\mathcal{H}$. Beginning with a state 
        $\ket{\Psi[\vec{a}(t=0)]}$ in $\mathcal{M}$, where $\vec{a}(t)$ are the 
        variational parameters, we wish to compute the time-evolution according 
        to the Schrödinger equation $\frac{\mathrm{d}}{\mathrm{d}t} 
        \ket{\Psi[\vec{a}(t)]} = -\mathrm{i} H \ket{\Psi[\vec{a}(t)]}$.
         
        The exact evolution generally leads out of $\mathcal{M}$. 
        Equivalently, the infinitesimal time step 
        $-\mathrm{i}H \ket{\Psi[\vec{a}(t)]}$ (the blue dashed arrow) need not lie 
        within the tangent plane $\mathbb{T}$ to $\mathcal{M}$ at point 
        $\ket{\Psi[\vec{a}(t)]}$ (the green dotted line). The best approximation 
        to the exact evolution, whilst remaining in $\mathcal{M}$, requires a 
        tangent vector $\ket{\Phi} \in \mathbb{T}$ (the red solid arrow) that best 
        approximates $-\mathrm{i}H \ket{\Psi[\vec{a}(t)]}$, which is found by
        projecting $-\mathrm{i}H \ket{\Psi[\vec{a}(t)]}$ onto $\mathbb{T}$. 
        In other words, $\ket{\Phi}$ must minimize 
        $\| \mathrm{i}H\ket{\Psi[\vec{a}(t)]} + \ket{\Phi}\|^2$.
        
        This is equivalent to finding optimal equations of motion for $\vec{a}$. 
        Writing $\ket{\Phi} =\dot{a}^j \ket{\partial_j\Psi}$ (where 
        $\ket{\partial_j\Psi} := \partial/{\partial a^j} \ket{\Psi[\vec{a}(t)]}$) 
        and taking the derivative of the above magnitude with respect to 
        $\dot{\overline{a}}^j$ results in the flow equations
        \begin{equation*}
            \mathrm{i} \dot{a}^j(t) = g^{jk}\braket{\overline{\partial_k} 
            \Psi| H | \Psi}
        \end{equation*}    
        where $g^{jk}$ is the inverse of $g_{jk} = \braket{\overline{\partial}_j 
        \Psi| \partial_k \Psi}$, which is the pullback metric on $\mathbb{T}$. 
        Here, we assume that $\ket{\Psi[\vec{a}(t)]}$ is a holomorphic 
        function of $\vec{a}(t)$, although this is not necessary.
      \end{minipage}
  }
  \label{TDVPbox}
\end{figure}

The simulation of infinite quantum spin systems has mostly been confined to the 
translation invariant setting (usually by restricting states to subsets of MPS 
that are either fully translation invariant or invariant under translations 
by $k > 1$ sites \cite{nebendahl_2013}). 
However, the ability to explore locally nonuniform 
states on an infinite lattice is particularly attractive for studying the 
dynamics, e.g.\ scattering, of localized excitations in large systems. For 
example, this would provide a realistic setting in which to study quantum field 
excitations. There has been some prior work in this direction, building on previous 
light-cone results \cite{osborne_2006, hastings_2009, banuls_2009}, where the 
dynamics of a local disturbance is (partially) studied in the Heisenberg picture. 
These approaches can become expensive for systems with large local spin dimensions 
(such as those appearing in lattice field theory). Another direction
that has been suggested \cite{kjall_2011},
is to work completely in the Schr\"odinger picture 
with infinite uniform MPS and to add a finite nonuniform region.

In this work we explore the locally optimal implementation of the TDVP for 
uniform MPS with a dynamically expandable nonuniform segment. We derive the 
equations of motion for the variational parameters using a particular choice of 
gauge-fixing which allows us to integrate the variational 
dynamics with a complexity that scales as $d|t|D^3 N$, where $N$ is the length 
of the nonuniform piece (the number of sites), $|t|$ is the desired integration 
time, $d$ is the local spin dimension, and $D$ is the bond dimension. Even 
though the ends of the nonuniform region can move, there may be some 
backscattering due to boundary effects; we describe how to compensate for these 
with the addition of an \emph{optical potential} term. These methods are 
illustrated in the case of local excitations of the spin-1 anti-ferromagnetic 
Heisenberg model and for particles in $\phi^4$ theory. 

We assume throughout that our Hamiltonian $H$ contains only nearest-neighbor 
terms. It is decomposed as $H = H^\text{uni} + H^\text{loc}$, where 
$H^\text{uni} = \sum_{n=-\infty}^\infty h^\text{uni}_{n,n+1}$ with 
$h^\text{uni}_{n,n+1} \cong h^\text{uni}_{m,m+1}$, $\forall n,m$, and 
$H^\text{loc} = \sum_{n=1}^{N-1} h^\text{loc}_{n,n+1}$ with $[1,N]$ representing 
a contiguous region of the lattice and $h^\text{loc}_{n,n+1} \equiv 0$ 
for $n < 1, n \ge N$, allowing us to also write $H = \sum_n h_{n,n+1} = 
\sum_n \left[ h^\text{uni}_{n,n+1} + h^\text{loc}_{n,n+1}\right]$. We consider 
two cases in particular: firstly, a non-trivial $h^\text{loc}$ leads to a 
locally nonuniform ground state, which can be found using imaginary time 
evolution via our algorithm. Secondly, given a purely uniform Hamiltonian 
($h^\text{loc}=0$) and an initial state that differs only locally (in a region 
$[1,N]$) from an eigenstate of $H^\text{uni}$, our algorithm can be used to 
simulate the resulting locally non-trivial dynamics.

To capture a locally nonuniform state using MPS, we define a class of 
``sandwich'' states (sMPS), based on uniform MPS, using two 
$d \times D \times D$ tensors $A_L$ and $A_R$ describing 
the (asymptotic) state either side of the nonuniform region $[1,N]$, which is 
parametrized by $N$ further tensors. An sMPS state can be written as
\begin{align*}
\ket{\Psi[A]} = \!\!\sum_{\{s\}=1}^{d} \!\!
                v_L^\dagger \! \left[\prod_{i=-\infty}^0 \!\!\! A_L^{s_i}\right] 
                \!\! A_1^{s_1} \dots A_N^{s_N} \!\!
                \left[\prod_{j=N+1}^\infty \!\!\! A_R^{s_j}\right] \!\! v_R
			    \ket{\vec{s}}
\end{align*}
where $\ket{\vec{s}} = \ket{\dots s_1 \dots s_N \dots}$ and 
$A^s_X \in M_D(\mathbb{C})$ (where $X = L,R,[1,N]$). Taking 
$A_L = A_1 = \dots = A_N = A_R$ gives a completely uniform state. The vectors 
$v_{L/R}$ are, as with uniform MPS \cite{haegeman_time-dependent_2011}, 
generically irrelevant to the TDVP algorithm and are not further specified. 
In principle, the dimensions of $A_X^s$ are subject only to the constraints of 
the matrix product, which can become important when maximizing numerical 
efficiency. However, for reasons of notational simplicity, we assume uniform 
dimensions here.

$A_{L/R}$ represent the left and right asymptotic states: the reduced density 
matrix $\rho_{[n,m]}(A_L, A_R, A_{1 \dots N})$ of a piece of the lattice in the 
left or right region $n,m < 1$ or $n,m > N$ tends to that of the uniform MPS 
state $\rho_{[n,m]}(A_{L/R})$ as the distance from the nonuniform region 
increases. Since $A_{L/R}$ represent infinite ``bulk'' regions of the lattice, 
their dynamics should not be affected by nonuniformities in the $[1,N]$ region,
which spread at a finite speed.
Furthermore, if the left and right asymptotic states are eigenstates of 
$H^\text{uni}$, they are left completely unchanged by time evolution. Assuming 
this, we restrict the variational parameters to the tensors $A_1\dots A_N$ 
and treat $A_{L/R}$ as boundary conditions. $A_{L/R}$ can be obtained for the 
ground state of $H^\text{uni}$ using the existing TDVP algorithm for uniform 
MPS \cite{haegeman_time-dependent_2011}. 
To accurately capture states with a nonuniform region $[1,N]$ in this way, 
$N$ should be sufficiently large so that the asymptotic states are already 
reached at the left and right boundaries with the bulk.

The tensor network formed by the matrices $A$ can be visualized as
\begin{center}
  \includegraphics{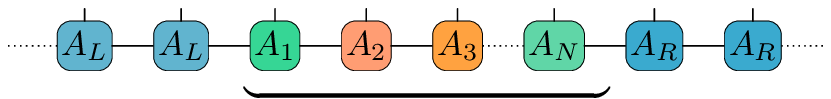}
\end{center}
with the nonuniform region marked in the center and the physical indices 
pointing upwards. 
Expectation values of local operators can be calculated efficiently in terms
of operators $E^{A}_{B} \equiv \sum_s^d A^s \otimes \overline{B}^s$, 
with the ``transfer operators'' $E_n \equiv E^{A_n}_{A_n}$.
For example, the expectation value of an operator $h$ that acts non-trivially on a 
pair of neighboring sites can be written as
\begin{align} \label{eqn:expval_long}
  \braket{h_{n,n+1}} = \braket{v_L|\left[ \prod_{k=-\infty}^{n-1} E_n \right] 
                                            E_n^h
                         \left[ \prod_{k=n+2}^{\infty} E_{n} \right] |v_R},
\end{align}
with $\bra{v_L} = v_L^\dagger \otimes \overline{v}_L^\dagger$ and
$\ket{v_R} = v_R \otimes \overline{v}_R$ as well as
$E_{n < 1} \equiv E_L$ and $E_{n > N} \equiv E_R$ and where
$E_n^h = \sum_{s,t,u,v}^d \braket{s,t|h|u,v} A_n^u A_{n+1}^v \otimes \overline{A_n^s A_{n+1}^t}$.
\begin{center}
  \includegraphics{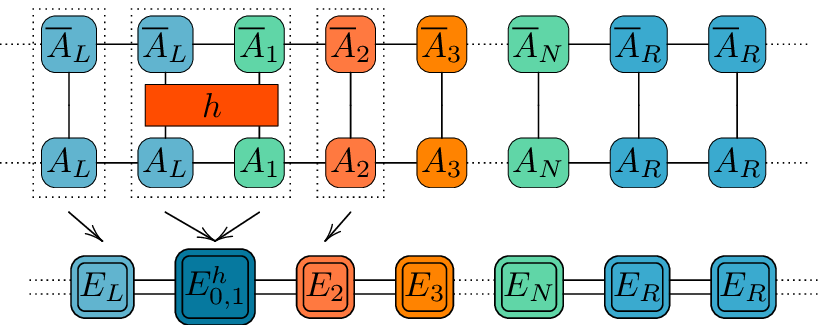}
\end{center}
Expressions for expectation values and for the norm of the state contain parts 
``$\bra{v_L}(E_L)^\infty$" and ``$(E_R)^\infty \ket{v_R}$" that need not
be well-defined, depending on the properties of $E_L$ and $E_R$.
To make these quantities finite, we must require that $E_{L/R}$ have spectral
radius equal to 1. To ensure that $v_L$ and $v_R$ remain irrelevant in 
calculations of bulk properties, we further demand that there is a single, 
non-degenerate (so that $A^s_{L/R}$ are not block diagonalizable)
eigenvalue of largest magnitude that is equal to 1, with all other eigenvalues
having magnitude strictly less that 1 \cite{haegeman_geometry_2012}.
The left and right eigenvectors corresponding to this eigenvalue, which
are thus the unique left and right fixed points of $E_{L/R}$, we name
$\bra{l_{L/R}}$ and $\ket{r_{L/R}}$, normalizing them such that 
$\braket{l_{L/R}|r_{L/R}} = 1$.
We can then write $\bra{x}(E_L)^\infty \propto \bra{l_L}$ and 
$(E_R)^\infty\ket{x} \propto \ket{r_R}$, where $\ket{x}$ is some vector that is not
orthogonal to $\bra{l_L}$ or $\ket{r_R}$. 

We now have a slightly simpler form for \eqref{eqn:expval_long}: 
$\braket{h_{n,n+1}} = \braket{l_L|\left[ \prod_{k=1}^{n-1} E_n \right] E_n^h
\left[ \prod_{k=n+2}^{N} E_{n} \right] |r_R}$.
To further improve the notation, we define $\bra{l_{n \ge 1}} = \bra{l_{n-1}} E_n$
and $\ket{r_{n < N}} = E_{n+1} \ket{r_{n+1}}$, identifying $\bra{l_{n<1}} \equiv \bra{l_L}$ 
and $\ket{r_{n\ge N}} \equiv \ket{r_R}$ (we will also use 
 $A_{n > N} \equiv A_R$ and $A_{n < 1} \equiv A_L$). 
We then have $\braket{h_{n,n+1}} = \rbra{l_{n-1}} E_n^{h} \rket{r_{n + 1}}$:
\begin{center}
  \includegraphics{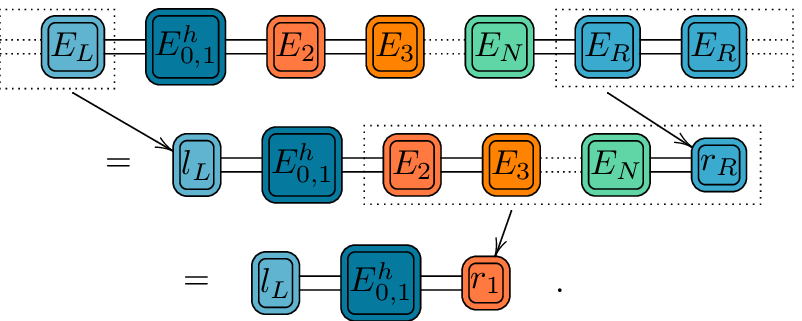}
\end{center}
Note that we are free to scale $\bra{l_L}$, $\ket{r_R}$ and the tensors $A_n$ of the
nonuniform region such that $\braket{\Psi|\Psi} = \braket{l_n|r_n} = 1, \forall n$.

For reasons of efficiency, when constructing numerical algorithms
we work in the isomorphic setting where transfer operators
are replaced by maps and vectors by matrices using the Choi-Jamiolkowski isomorphism.
Here, a $D^2 \times D^2$ transfer operator acting on a vector $E_n\ket{x}$ becomes 
$\sum_s^d A_n^s x {A_n^s}^\dagger$ with $x$ a $D \times D$ matrix, 
so that expectation values can be computed
using $\mathcal{O}(D^3)$ scalar multiplication operations:
\begin{center}
  \includegraphics{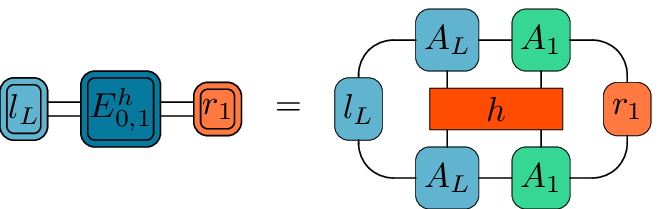}.
\end{center}

We now determine the dimension of the sub-manifold $\mathcal{M}_\text{sMPS} \subset \mathcal{H}$ 
of Hilbert space defined by the sMPS variational class. Naively, this is the 
number of complex entries of the parameter tensors $A_{1\dots N}$, which is $N d D^2$. 
However, an sMPS state is invariant under gauge transformations
\begin{align}
 \label{eqn:gt}
 \begin{split}
   A^s_L &\rightarrow g_0 A^s_L g_0^{-1}  \\
   A_{1\le n \le N}^s &\rightarrow g_{n-1} A_n^s g_n^{-1} \\
   A^s_R &\rightarrow g_N A^s_R g_N^{-1}
 \end{split}
\end{align}
with $g_n \in M_D(\mathbb{C})$. 
Since $A_{L/R}$ are fixed, we restrict to $g_0 = g_N = \mathbb{I}$
leaving $(N - 1)D^2$ non-physical degrees of freedom corresponding to the 
gauge-transformation matrices $g_{1\dots (N-1)}$, as well as a further
one corresponding to the norm and phase.
The dimension of the sMPS variational manifold is thus
$\dim(\mathcal{M}_\text{sMPS}) = (N(d-1) + 1)D^2 - 1$. The redundancy in the
sMPS representation is familiar from other MPS variational classes 
\cite{haegeman_time-dependent_2011} and is less inconvenient than it may appear,
since the gauge-freedom in the representation of tangent vectors allows for
significant simplification of the TDVP flow equations.

To implement the TDVP (see boxout), we must project exact infinitesimal
time steps $-\ic H \ket{\Psi[A]}$ onto the tangent plane $\mathbb{T}_{\ket{\Psi[A]}}$ 
to $\mathcal{M}_\text{sMPS}$ at the point $\ket{\Psi[A]}$. The tangent plane is 
spanned by tangent vectors 
\begin{align}
  \label{eqn:tangvec}
  &\ket{\Phi[B]} = \sum_{n=1}^{N} \sum_{i=1}^{dD^2} B_{n,i} \ket{\partial_{n,i}\Psi[A]} \\
  &\quad\includegraphics{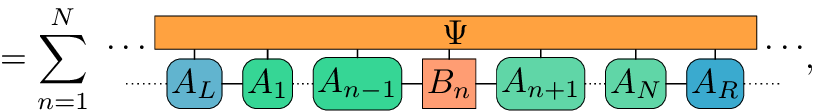} \nonumber
\end{align}
with $\ket{\partial_{n,i}\Psi[A]} = \partial/\partial A_{n,i} \ket{\Psi[A]}$
and the index $i$ running over all $dD^2$ entries of each tensor $A_n$ or $B_n$. 
The projection is achieved by finding a $\ket{\Phi[B]}$ that satisfies 
\begin{align}
  \label{eqn:tdvp_mps}
  \ket{\Phi[B]} = \arg \min_{\ket{\Phi[B']}}\| \mathrm{i}H\ket{\Psi[A(t)]} + \ket{\Phi[B']}\|^2.
\end{align}
Expanding the RHS leaves terms $\braket{\Phi[\overline{B}]|\Phi[B]}$ and 
$\braket{\Phi[\overline{B}] | H | \Psi[A]} + \text{h.c.}$, where the remaining 
$H^2$ term is a constant that can be ignored. The metric term
$\braket{\Phi[\overline{B}]|\Phi[B]}$ is at first glance very
complicated, since it couples the tensors $B_n$ for different lattice sites
in terms such as
\begin{align}
  \label{eqn:metric_term_ex}
  \sum_{n < m} \; \vcenter{\hbox{\includegraphics{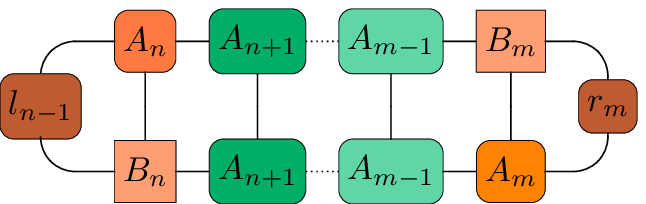}}}\;,
\end{align}   
precluding a splitting of the problem into $N$ separate parts (one for each 
$B_n$). Fortunately, these site-mixing terms can be eliminated by fixing the
gauge-freedom in the tangent vector representation.
If we impose the left gauge-fixing conditions (GFC) 
\begin{align}
  \label{eqn:gfc1}
  \begin{split}
    \rbra{l_{n-1}} E^{B_n}_{A_n} &= 0 = \sum_s^d {A_{n}^s}^\dagger l_{n-1} B_{n}^s \\
    \vcenter{\hbox{\includegraphics{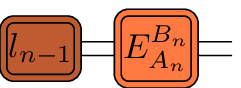}}} &= 0 =
    \vcenter{\hbox{\includegraphics{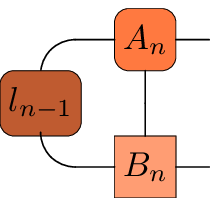}}}
  \end{split}
\end{align}
for sites $1 \le n < n_c$ and the right gauge-fixing conditions
\begin{align}
  \label{eqn:gfc2}
  \begin{split}
    E^{B_n}_{A_n} \rket{r_n} &= 0 = \sum_s^d B_{n}^s r_n {A_{n}^s}^\dagger \\
    \vcenter{\hbox{\includegraphics{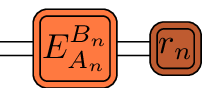}}} &= 0 =
    \vcenter{\hbox{\includegraphics{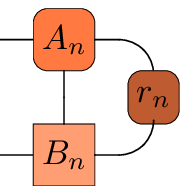}}}
  \end{split}
\end{align}
for sites $n_c < n \le N$,
we eliminate all site-mixing terms like \eqref{eqn:metric_term_ex} such that
$\braket{\Phi[\overline{B}]|\Phi[B]} = \sum_{n=1}^N \rbra{l_{n-1}} 
E^{B_n}_{B_n} \rket{r_n}$. 
Note that, for some site $n_c$ in the nonuniform region, 
the tangent vector parameters $B_{n_c}$ are not constrained. 
For reasons of symmetry, we choose $n_c$ to be in the middle 
so that $2 (n_c - 1) + 1 = N$ with odd $N$.

To see that the conditions \eqref{eqn:gfc1} and \eqref{eqn:gfc2} 
fix exactly the gauge degrees of freedom, 
we consider the one-parameter gauge transformation 
$g_n(\eta) = \mathbb{I} + \eta x_n \quad \forall n \in [0\dots N]$ with
$x_0 = x_N = 0$. Writing the transformed state as $\ket{\Psi[A'(\eta)]}$,
the infinitesimal transformation has the form \eqref{eqn:tangvec} of a tangent vector
\begin{align*}
  \left. \frac{\mathrm{d}}{\mathrm{d} \eta} \ket{\Psi[A'(\eta)]} \right|_{\eta = 0} = \ket{\Phi[\mathcal{N}[x]]} = 0,
\end{align*}
with $\mathcal{N}^s_n[x] = x_{n-1} A_n^s - A_n^s x_n$. Tangent vector parameters 
of this form thus capture exactly the gauge freedom so that an arbitrary tangent vector
fulfills $\ket{\Phi[B]} = \ket{\Phi[B + \mathcal{N}[x]]}$. Using this freedom,
we can always transform arbitrary $B'_n$ as $B_n = B'_n + \mathcal{N}_n[x]$ so that
$B_n$ satisfies the gauge-fixing conditions \eqref{eqn:gfc1} and \eqref{eqn:gfc2}. 
To see this, we insert $B'_n + \mathcal{N}_n[x]$ into \eqref{eqn:gfc2} to obtain
\begin{align*}
  x_{n-1} r_{n-1} = \sum_s^d (A_n^s x_n - {B'_n}^s) r_n {A_n^s}^\dagger \quad \forall n > n_c,
\end{align*}
which we can solve to fully determine $x_{n-1}$ given that $r_{n-1}$ has full rank 
and that $x_n$ is known. Starting at $n=N$ with $x_N = 0$, this fixes all $x_n$ down
to $n=n_c$. We can perform the same trick with \eqref{eqn:gfc1} to get
\begin{align*}
     l_n x_n = \sum_s^d {A_n^s}^\dagger l_{n-1} (x_{n-1} A_n^s + {B'_n}^s) \quad \forall n < n_c,
\end{align*}
which determines the remaining $x_n$ (up to $n=n_c - 1$) 
given that $x_0 = 0$ and that $l_n$ has full rank.

We can construct $B_n$ such that they automatically fulfill the GFC \eqref{eqn:gfc1} and \eqref{eqn:gfc2}. 
For $n_c < n \le N$ we define the $(d - 1)D \times dD$ 
matrix $V_n^\dagger$ to contain an orthonormal basis for the null space of 
$[R_n]_{(\alpha, s); \beta} = [r_n^{1/2} {A_n^s}^\dagger]_{\alpha, \beta}$ and
set 
\begin{align}
  \label{eqn:param2}
  B_{n}^s(x_n) = l_{n-1}^{-1/2} x_n V_n^s r_n^{-1/2} \quad \forall n \in [n_c + 1, N],
\end{align}
with parameters $x_n$.
For $1 \le n < n_c$, we define the $dD \times (d - 1)D$ matrix $W_n$
to contain an orthonormal basis for the null space of 
$[L_n]_{\alpha; (s, \beta)} = [{A_n^s}^\dagger l_{n-1}^{1/2}]_{\alpha,\beta}$
and set
\begin{align}
  \label{eqn:param1}
  B_{n}^s(x_n) = l_{n-1}^{-1/2} W_n^s x_n r_n^{-1/2} \quad \forall n \in [1, n_c - 1].
\end{align}
It is easy to check by insertion that \eqref{eqn:param1} and \eqref{eqn:param2} 
respectively satisfy the GFC \eqref{eqn:gfc1} and \eqref{eqn:gfc2}.
Note again that $B_{n_c}$ remains unconstrained.
Using the parametrizations, we obtain 
\begin{align}
  \label{eqn:metric_term_param}
  \braket{\Phi[\overline{B}]|\Phi[B]} = 
  \sum_{n\neq n_c} \tr[x_n^\dagger x_n] + \braket{l_{n_c - 1}|E^{B_{n_c}}_{B_{n_c}}|r_{n_c}}.
\end{align} 

Having fixed the gauge, one non-physical degree of freedom remains, since
$\braket{\Psi[\overline{A}]|\Phi[B]} = \braket{l_{n_c - 1}|E^{B_{n_c}}_{A_{n_c}}| r_{n_c}} \neq 0$, 
implying that the tangent plane
contains infinitesimal changes to the norm and phase. We must thus explicitly
eliminate norm and phase changes when implementing the TDVP,
which can be done by replacing $H$ 
with $\tilde{H} \equiv H - \braket{\Psi|H|\Psi}$ in the TDVP flow equations, 
effectively projecting out the corresponding component of $H\ket{\Psi}$ 
\cite{haegeman_time-dependent_2011}.

With gauge-fixing, $\braket{\Phi[\overline{B}] | \tilde{H} | \Psi[A]}$
simplifies, but still contains terms mixing $B_n$ and $\tilde{h}_{m,m+1} \equiv 
h_{m,m+1} - \braket{h_{m,m+1}}$ for 
\smash{$m,m+1 \neq n$}. Each $B_n$ term contains a sum over $\tilde{h}_{m, m+1}$
extending into the left ($n < n_c$) or right ($n > n_c$) bulk or into both ($n = n_c$).
This is understood by defining the right and left effective Hamiltonians 
\begin{align*}
  \rket{{K}_{n+1}} = \sum_{m=n+1}^\infty \left( \prod_{k=n+1}^{m-1} E_k \right) E^{\tilde{h}}_{m} \rket{r_{m+1}} \quad \text{and} \\
  \rbra{{J}_{n-1}} = \sum_{m=-\infty}^{n-2} \rbra{l_{m-1}} E^{\tilde{h}}_{m} \left( \prod_{k=m+2}^{n-1} E_k \right),
\end{align*}
which also obey
\begin{align*}
  \rket{{K}_{n}} &= E_{n}\rket{{K}_{n+1}} + E^{\tilde{h}}_{n}\rket{r_{n+1}} \quad \text{and} \\
  \rbra{{J}_{n}} &= \rbra{{J}_{n-1}} E_{n} + \rbra{l_{n-2}} E^{\tilde{h}}_{n-1},
\end{align*} 
where $E^{\tilde{h}}_{n} = \sum_{s,t=1}^d {C}_n^{s,t} \otimes 
\overline{A_n^s A_{n+1}^t}$ and
${C}_n^{s,t} = \sum_{u,v=1}^{d} \braket{s,t|\tilde{h}_{n,n+1}|u,v} 
A_n^u A_{n+1}^v$. For example, the terms containing $B_n$ with $n > n_c$ are:
\begin{align*}
  \sum_{n_c < n < m}& \vcenter{\hbox{\includegraphics{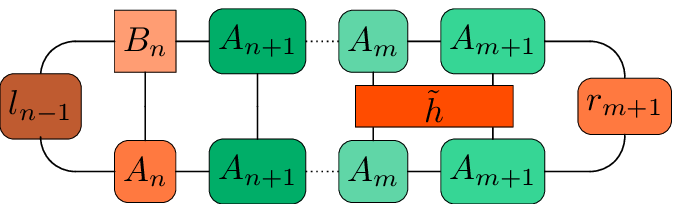}}} \\
   &= \sum_{n_c < n} \;\vcenter{\hbox{\includegraphics{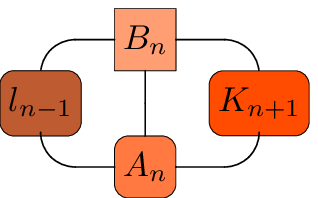}}}\;.
\end{align*}
The sums over the uniform bulk $\rket{{K}_R} 
\equiv \rket{{K}_{N+1}} = \sum_{n=0}^\infty (E_R)^n E_R^{\tilde{h}} \rket{r_R}$
 and $\rbra{{J}_L} \equiv \rbra{{J}_{0}} = \sum_{n=0}^\infty \rbra{l_L} E_L^{\tilde{h}} (E_L)^n$
can be computed by exploiting the assumption that 
$E_{L/R}$ have a unique largest (in magnitude) eigenvalue equal to 1,
which allows us to rewrite the sum as a pseudo-inverse.
For the right-hand bulk this gives 
$\rket{{K}_R} = (\mathbb{I} - E_R)^\text{P} E_R^{\tilde{h}} 
\rket{r_R}$ or, equivalently, \mbox{$(\mathbb{I} - (E_R - 
\rket{r_R}\rbra{l_R}))\rket{{K}_R} = E_R^{\tilde{h}} \rket{r_R}$}, which can 
then be solved for ${K}_R$ in the matrix representation 
using $\mathcal{O}(D^3)$ operations per iteration. 
$\rbra{J_L}$ can be computed analogously.
Note that the energy difference due to the nonuniformity is 
$\Delta E = \braket{{J}_L|r_0} + \braket{l_N|{K}_R} + 
\sum_{n=0}^N \braket{h_{n, n+1}} - (N+1)\braket{h}_\text{uni}$, 
where $\braket{h}_\text{uni}$ is the energy per-site of the uniform bulk state. 

We now have the ingredients needed to compute the Hamiltonian term efficiently as
\begin{align}
  \braket{\Phi[\overline{B}]|\tilde{H}|\Psi[A]}& \label{eqn:ham_term} \\
   = \sum_{n \neq n_c} &\tr\left[ x_n^\dagger F_n \right] + \sum_{s=1}^d \tr\left[ 
  l_{n_c-1} G_{n_c}^s r_{n_c} {B_{n_c}^s}^\dagger \right], \nonumber
\end{align}
with
\begin{align}
  F_{n > n_c} &\equiv \sum_{s,t}^d l_{n-1}^{1/2} {C}_n^{s,t} r_{n+1} {A^t_{n+1}}^\dagger r_n^{-1/2} {V^s_n}^\dagger \nonumber \\
  &+ \sum_{s,t}^d l_{n-1}^{-1/2} {A^t_{n-1}}^\dagger l_{n-2} {C}_n^{t,s} r_n^{1/2} {V^s_n}^\dagger \nonumber\\
  &+ \sum_{s}^d l_{n-1}^{1/2} {A^s_{n}} {K}_{n+1} r_n^{-1/2} {V^s_n}^\dagger, \nonumber
\end{align}
\begin{align}
  F_{n < n_c} &\equiv \sum_{s,t}^d {W_n^s}^\dagger l_{n-1}^{1/2} C_n^{s,t} r_{n+1} {A_{n+1}^t}^\dagger r_n^{-1/2} \nonumber \\
  &+ \sum_{s,t}^d {W_n^s}^\dagger l_{n-1}^{-1/2} {A_{n-1}^t}^\dagger l_{n-2} C_{n-1}^{t,s} r_n^{1/2} \nonumber\\
  &+ \sum_{s}^d {W_n^s}^\dagger l_{n-1}^{-1/2} J_{n-1} A_n^s r_n^{1/2},\quad\text{and} \nonumber
\end{align}
\begin{align}
  G_{n_c}^s &\equiv A_{n_c}^s {K}_{n_c + 1} r_{n_c}^{-1} + l_{n_c - 1}^{-1} {J}_{n_c - 1} A_{n_c}^s \nonumber\\
  &+ \sum_{t}^d \left[ {C}_{n_c}^{s,t} r_{n_c+1} {A_{n_c+1}^t}^\dagger r_{n_c}^{-1} \right. \nonumber \\
  & \left.\qquad + l_{n_c-1}^{-1} {A_{n_c-1}^t}^\dagger l_{n_c - 2} {C}_{n_c -1}^{t,s} \right], \nonumber
\end{align}
where $n \in [1, N]$ and $J_n$ is the conjugate matrix representation of $\bra{J_n}$ so that, 
for some vector $\ket{y}$, $\braket{J_n|y} = \tr[J_n y]$.

Having fixed the gauge, inserting \eqref{eqn:metric_term_param} and 
\eqref{eqn:ham_term} into the TDVP minimization problem 
\eqref{eqn:tdvp_mps} and minimizing over the parameters $x_{n \neq n_c}$
and $B_{n_c}$ gives us $N - 1 + d$ independent matrix equations,
\begin{align*}
  B_{n_c}^s = -\ic G_{n_c}^s \;\; (s \in [1,d]) \quad \text{and} \quad x_n = -\ic F_n \;\;  (n \neq n_c),
\end{align*}
representing the optimal time evolution for the variational parameters 
\begin{align}
  \label{eqn:tdvp_flow_mps}
  \dot{A}_{n_c}^s = -\mathrm{i}G_{n_c}^s(A) \quad \text{and} \quad
  \dot{A}^s_{n \neq n_c} = B^s_n(-\mathrm{i}F_n(A)),
\end{align}
where we use the appropriate parametrization \eqref{eqn:param2} or \eqref{eqn:param1}
for $B_n$ depending on the value of $n$. 
With gauge-fixing, the independent terms 
to be minimized in \eqref{eqn:tdvp_mps}, one for each $B_n$, can be summarized 
diagrammatically as
\begin{widetext}
$
  \vcenter{\hbox{\includegraphics{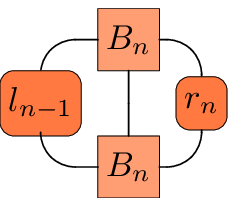}}} \;=\; 
  \vcenter{\hbox{\includegraphics{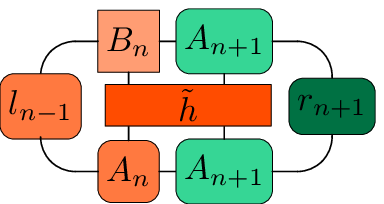}}} 
  \;+\;\,\vcenter{\hbox{\includegraphics{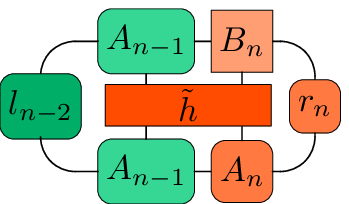}}} 
  \;+\; \underbrace{\vcenter{\hbox{\includegraphics{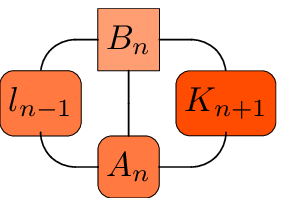}}}}_{\text{only for } n \ge n_c}
  \;+\; \underbrace{\vcenter{\hbox{\includegraphics{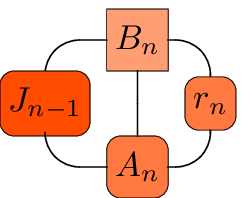}}}}_{\text{only for } n \le n_c} \,,
$
\end{widetext}
where the equations for $x_n$ are obtained again by replacing $B_n$ with \eqref{eqn:param2} or \eqref{eqn:param1} for
$n \neq n_c$ as appropriate.
The flow equations \eqref{eqn:tdvp_flow_mps} can be integrated numerically, for example 
with the following simple algorithm implementing the Euler method:
\begin{enumerate}
  \item Calculate $F_n, G_{n_c}^s$.
  \item Take a step by setting $A_n(t + dt) = A_n(t) + dt B_n$.
  \item Restore a canonical form using a gauge transformation \eqref{eqn:gt} and 
  normalize the state by rescaling $A_{n_c}$.
  \item Compute desired quantities, such as the energy expectation value, 
  and adjust the step size $dt$ as required.
  \item If needed, expand the nonuniform region to the left and/or right.
\end{enumerate}
Normalization is necessary because the norm is only preserved to first order in 
$dt$. Maintaining a canonical form (for example, see appendix \ref{sec:app_can_form}) 
can simplify some parts of the TDVP calculations and improve the conditioning of the matrices 
involved. The last step allows for a small initial nonuniform region, which can 
be grown if the dynamics warrant changing the state significantly outside of it. 
This is done by ``absorbing'' sites from the uniform region(s) into the 
nonuniform region, copying the $A_L$ and $A_R$ matrices as needed. 

Whether it is necessary to grow the nonuniform region can be heuristically 
determined by observing the per-site contributions 
$\eta_n = \sqrt{\braket{l_{n-1}|E^{B_n}_{B_n}|r_n}}$ to the norm 
$\eta = \sum_n \eta_n$ of the TDVP tangent vector $\ket{\Phi[B]}$.
If $\eta_1$ and $\eta_N$ become significantly larger than the norm $\eta_\text{uni}$ 
of the uniform MPS TDVP tangent vector of the bulk state then the
nonuniform region should be expanded until this is no longer the case.

Note also that the above algorithm is not well suited to simulating real-time 
dynamics because errors due to the simple integration method used are cumulative. 
Instead, more sophisticated integrators such as the commonly used fourth-order 
explicit Runge-Kutta method (see appendix \ref{sec:RK4}) are preferable. The Euler 
method is, however, still useful for finding ground states because imaginary time 
evolution is ``self-correcting'' | it will always take you towards the ground 
state, given that the starting point is not orthogonal to it.

\begin{figure}[t]
    \includegraphics{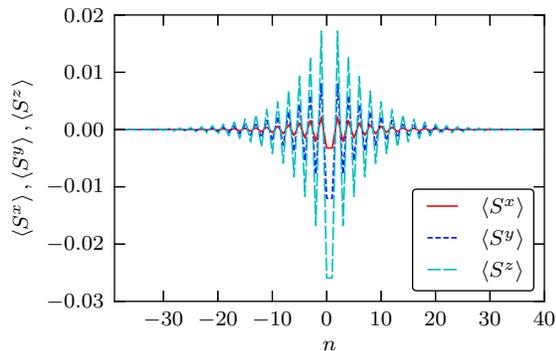}
    \caption{Per-site spin expectation values of the approximate ground
    state of the spin-1 antiferromagnetic Heisenberg model with a ferromagnetic 
    impurity. The state was converged up to $\eta \approx 3 \times 10^{-8}$ with
    $D = 64$ and a nonuniform region $[-100, 100]$. The initial uniform ground 
    state used for the left and right bulk
    parts was converged to $\eta \approx 10^{-12}$. (Color online.)
    }
    \label{fig:heis_imp}
\end{figure}

To test our algorithm, we use the antiferromagnetic spin-1 Heisenberg 
model $h^\text{uni}_{n,n+1} = h^\text{AFH}_{n,n+1}$, with
\begin{align}
  \label{eqn:spin1_AF_heis}
  h^\text{AFH}_{n,n+1} = \sum_{i=x,y,z} S^i_n S^i_{n+1}. 
\end{align}  
The uniform ground state respects the SU(2) symmetry of the Hamiltonian.
Having found a uniform MPS approximation for the ground state, we 
use imaginary time evolution to find the ground state of a nonuniform
model where one of the coupling terms has its sign flipped via the
addition of $h^\text{loc}_{0,1} = -2 h^\text{AFH}_{0,1}$, with all 
other $h^\text{loc}_{n\neq 0,n+1} = 0$, thus creating a ferromagnetic
impurity. 
Impurities have been studied in this model before 
\cite{kaburagi_1993, *kaburagi_1994, *sorensen_1995, *wang1_1996, *wang2_1996} 
however, to the best of our knowledge the case of a ferromagnetic bond has not yet been 
investigated. It appears to lead to localized SU(2) symmetry-breaking, as can be 
seen in the relative distribution of the 
spin expectation values at each site, which we plot in Fig.~\ref{fig:heis_imp}. 
This is expected, since the ground states of the
uniform ferromagnetic model also break the symmetry. In this case,
$-\sum_{i}S^i_0 S^i_1$ acts in the Hamiltonian to approximately 
project the pair of sites $0$ and $1$ onto the spin 2 subspace, 
whose states are not invariant under SU(2).

\begin{figure}[t]
    \includegraphics{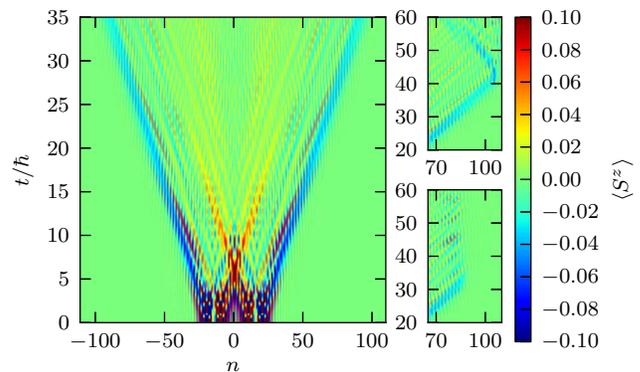}
    \caption{Simulated real time evolution (using a fixed nonuniform region 
    $[-100, 100]$ with $D = 64$ and $dt = 0.04$) of the spin-1 anti-ferromagnetic 
    Heisenberg model with two localized entangled excitations generated by 
    applying $S_{m-j}^-S_{m+j}^+$ at $m=\pm15$ with $j = \pm5$. The plots show 
    the expectation value of $S^z_n$ with the top-right plot showing the 
    excitation bouncing at the right boundary. For the bottom-right plot, we 
    used a Gaussian optical potential to suppress this reflection, albeit 
    imperfectly, with $\epsilon_n = e^{-(n + 90)^2 / 18 } + e^{-(n - 90)^2 / 18 }$. 
    The uniform ground state was converged up to a state tolerance $\eta = 10^{-8}$. 
    For the time-integration we used a 4th order explicit Runge-Kutta algorithm.
    (Color online.)}
    \label{heis_real}
\end{figure}

\begin{figure}[t]
    \includegraphics{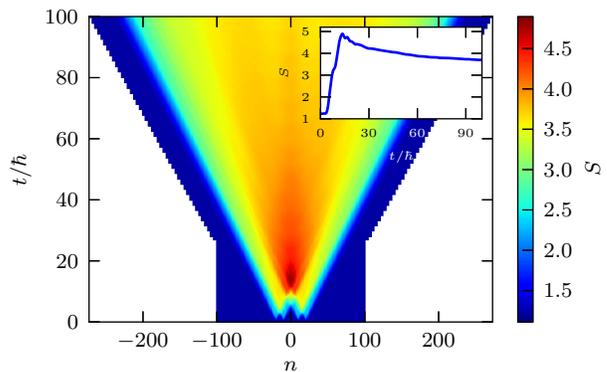}
    \caption{
        The time evolution of the block entropy $S$ of one half of the lattice, 
        split at each site $n$, for the same simulation as in Fig.~\ref{heis_real},
        except that dynamic expansion of the nonuniform region is used, as
        indicated by the ``staircase'' pattern | the bulk parts are shown in white. 
        The inset shows a cross-section at site $0$. (Color online.)
    }
    \label{heis_entr}
\end{figure}

As a test of real-time evolution, we again use $h^\text{uni} = h^\text{AFH}_{n,n+1}$ 
from \eqref{eqn:spin1_AF_heis}, but without any local perturbations 
($h^\text{loc}_{n,n+1} = 0$, $\forall n$). We begin with a uniform ground state
approximation and introduce local 
excitations by applying the (nonunitary) operator 
$S_{m-j}^-S_{m+j}^+$ with $S_n^\pm = S^x \pm \ic S^y$, 
which generates an entangled excitation,
to two separated pairs of sites at $m = \pm k$ inside a nonuniform region. 
By calculating the expectation value of an observable such as $S^z$ for a 
set of sites (possibly extending into the left 
and right bulk regions) after each step, the time evolution of the system can 
be visualized, for example by plotting the site spin expectation values as in 
Fig.~\ref{heis_real} or the half-chain entropy for splittings at each site as
in Fig.~\ref{heis_entr}. For the latter, we use dynamic expansion
of the nonuniform region to maximize numerical efficiency.
Note that the entropy for a splitting after site $0$ appears to tend to an 
asymptotic value of approximately $3.5$. This suggests that a hybrid method whereby uniform matrices
are reintroduced between the two excitations as they become separated could be used 
to study the asymptotics of entangled excitations for large times.

To mitigate non-physical reflections
that can occur when a traveling excitation meets a boundary with the uniform 
region, ``optical potential'' terms $h^\text{loc}_{n,n+1} = 
-\mathrm{i}\epsilon_n h^\text{uni}_{n,n+1}$ can be locally turned on near to the 
boundaries. This effectively carries out imaginary time evolution on a subsystem
defined by the envelope function $\epsilon(n)$, where the magnitude of $\epsilon(n)$
determines the rate of ``cooling'' at each site. If $\epsilon(n)$ is a step function
that turns on imaginary time evolution at a constant rate in a small part of the
lattice, that part should (in the absence of simultaneous real time evolution) 
converge to the ground state of a finite chain with open boundary conditions. 
Since we are working with gapped systems, the ground state of a 
smaller part should be the same as that of the uniform infinite system up to 
boundary effects.
We find that choosing $\epsilon(n)$ to be superposition of two gaussians, each
localized near an edge of the nonuniform region, avoids
significant boundary effects during evolution of the Heisenberg model \eqref{eqn:spin1_AF_heis}
whilst successfully attenuating boundary reflections, as shown in 
Fig.~\ref{heis_real}. Note that the entanglement present in the 
excitations produced for this particular model mean that the boundary-absorption
affects the evolution in the central region as well as at the boundaries themselves.
Further tuning of $\epsilon_n$ may help to more effectively dissipate 
the excitations heading out of the nonuniform region. 

\begin{figure}[t] 
    \includegraphics{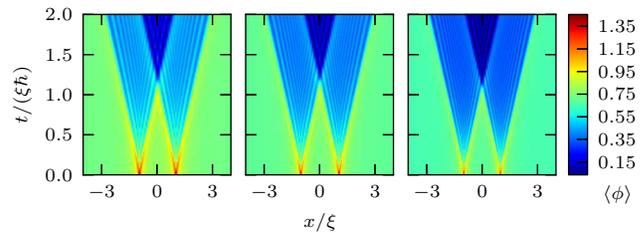}
    \caption{Simulation ($D=24$, $d=16$, $dt=0.02$) of particle scattering 
    in $(1+1)$-dimensional lattice $\phi^4$ real scalar field theory approaching 
    a continuum limit (from left to right), as determined by the ground-state
    correlation length in lattice sites $\tilde \xi$.
    We created two excitations by applying the field operator $\phi_n$ at two different
    sites $n = \pm [\tilde \xi]$ to an approximate ground state
    in the symmetry-broken phase. The coupling is $\tilde \lambda = 0.2$ for all 
    three plots and the parameter ratio $\tilde \lambda / \tilde \mu_R^2$
    varies from left to right as $\tilde \lambda / \tilde \mu_R^2 = 85, 80, 75$.
    Distance $x$ and time $t$ are scaled with the correlation length $\xi = a \tilde \xi$
    where $a$ is the lattice spacing. The field expectation value $\braket{\phi}$
    is left unscaled, although a more comprehensive treatment would scale it with the 
    field strength renormalization factor, which can also be computed from the
    uniform MPS approximate ground state.
    (Color online.)
    }
    \label{phi4}
\end{figure}

As a final test of our approach we simulate the scattering of localized
excitations in $\phi^4$ theory on a one-dimensional lattice.
The Hamiltonian is 
\begin{align}
  \label{eqn:ham_phi4}
  H^{\phi^4} = \sum_n \left(\frac{1}{2} \pi_n^2 + 
    \frac{\tilde \mu_0^2}{2} \phi_n^2 + \frac{\tilde \lambda}{4!} \phi_n^4 + \frac{1}{2} 
    (\phi_n - \phi_{n+1})^2 \right), 
\end{align}
where $[\phi_n(t), \pi_m(t)] = 
\mathrm{i}\delta_{nm}$. The bare mass $\tilde \mu_0$ and 
coupling $\tilde \lambda$ are dimensionless lattice parameters
related to parameters with dimension $[\text{mass}]^2$ by 
$\tilde \mu_0^2 = a^2 \mu_0^2$, $\tilde \lambda = a^2 \lambda$, where
$a$ is the lattice spacing. We fix $a$ for each set of parameters using
the ground state correlation length in lattice sites $\tilde \xi$, which 
is directly obtainable\cite{haegeman_post-matrix_2013} from the largest two 
eigenvalues of the uniform MPS transfer operator $E$.
Due to renormalization, $\tilde \mu_0$ is not equal to the particle mass and
in fact diverges in the continuum limit. 
So that our parameters are well-defined in the limit, we
separate out the divergent contribution $\delta \tilde \mu^2$ 
to obtain the renormalized mass-squared parameter
$\tilde \mu_R^2 = \tilde \mu_0^2 - \delta \tilde \mu^2$.
For certain values of $\tilde \mu_R^2, \tilde \lambda$ the ground state 
spontaneously breaks the global $\mathbb{Z}_2$ symmetry $\phi_n = -\phi_n$
of \eqref{eqn:ham_phi4} such that 
$\braket{\phi_n} = \pm \phi_0$.
In Fig.~\ref{phi4}, we examine excitations of $\phi^4$ theory within a 
nonuniform region by applying the field operator
to the ground state and simulating time-evolution. We do this for a sequence
of parameters, approaching a continuum limit.
More details about the application of MPS to 
real scalar $\phi^4$ theory and its critical behavior are available 
elsewhere \cite{sugihara_density_2004, us_upcoming}.

In this paper, we have introduced an efficient means of simulating the dynamics 
of localized nonuniformities on spin chains in the thermodynamic limit using the 
time-dependent variational principle (TDVP) and a special class of matrix 
product states (MPS). As with the existing algorithms implementing the TDVP for 
MPS in other settings \cite{haegeman_time-dependent_2011}, this algorithm 
approximates exact time evolution optimally given the restrictions of the 
variational class. Our (open source) implementation evoMPS \cite{evoMPS} 
is available as Python (http://www.python.org) source code, including example simulation 
scripts.

During completion of this work, we learned of other independent results
\cite{phien_infinite_2012, phien_dynamical_2012, zauner_time_2012} that use 
time-evolving block decimation to approximate the time evolution of a nonuniform
window on an otherwise translation-invariant chain. Our approach differs in that
we define a variational class and apply the TDVP to obtain equations for locally 
optimal approximate time evolution. We are then able to apply standard numerical
integration techniques. The idea of not only growing the nonuniform region,
but also of ignoring the evolution of uninteresting parts of the nonuniformity
for reasons of efficiency | say, to follow a wavefront \cite{phien_dynamical_2012, zauner_time_2012} 
can also be implemented in our scheme by restricting the variational parameters 
to a smaller part of the nonuniform region and leaving the rest constant (up to gauge
transformations). As mentioned above, another approach to studying
entangled excitations may be to detect when the central region between two separating
wavefronts becomes translation invariant over a sufficiently large region, 
taking this state as a new bulk state for one side of the system and restricting the
nonuniform region to a single wavefront.

\emph{Acknowledgements} | Helpful discussions with Florian Richter, Fabian Transchel
and Fabian Furrer are gratefully acknowledged.
This work was supported by the ERC grants QFTCMPS, QUERG and QUEVADIS, the FWF SFB grants
FoQuS and ViCoM and the cluster of excellence EXC 201 Quantum Engineering and Space-Time
Research.

\appendix

\section{Canonical form}
\label{sec:app_can_form}

A canonical form that fits to the gauge-fixing conditions (GFC) \eqref{eqn:gfc1} and 
\eqref{eqn:gfc2} is given by
\begin{align*}
  l_n &= \mathbb{I} \quad \forall n < n_c, \\
  l_n &= \mathrm{diag}(\lambda_{n,1}^2 \dots \lambda_{n,D}^2) \quad \forall n \in [n_c, N],\\
  r_n &= \mathbb{I} \quad \forall n \ge n_c,\\
  r_n &= \mathrm{diag}(\lambda_{n,1}^2 \dots \lambda_{n,D}^2) \quad \forall n \in [0, n_c - 1],
\end{align*}
where $\lambda_{n,i}$ for $1 \le i \le D$ are the Schmidt coefficients for the decomposition 
of the chain into two infinite halves by cutting between sites $n$ and $n+1$. It
corresponds to the GFC in the sense that changing the parameters as 
$A_n \rightarrow A_n + \epsilon B_n$ with $B_n$ satisfying the GFC does not alter $l_{n<n_c}$
or $r_{n \ge n_c}$, which are constants in the above canonical form, to first order in $\epsilon$.
In practice, this means that the canonical form is approximately maintained when making
finite steps in the TDVP algorithm.

The above canonical form can be reached via a gauge-transformation $g_{0 \dots N}$ where
$g_0$ and $g_N$ are non-trivial (see \eqref{eqn:gt}), such that the uniform bulk parameters $A_{L/R}$ are
also transformed. Since the overall state and also the left and right uniform bulk states
are unaffected by these transformations, performing them does not affect evolution under 
the TDVP equations.

\section{Runge-Kutta integration}
\label{sec:RK4}
For real-time evolution, numerical integration using the Euler method is
inefficient since small step sizes $\mathrm{d}t$ are required to keep
the $\mathcal{O}(\mathrm{d}t^2)$ integration errors made with each finite 
step small. A well known integration method with more favorable error
scaling is the 4th order Runge-Kutta method (RK4) \cite{press_section_????}, 
which makes per-step errors 
$\mathcal{O}(\mathrm{d}t^5)$ at the cost of three extra evaluations of the
derivative. It builds a final step by making three smaller steps and weighting
the derivatives obtained at the visited points.
Given a differential equation $\dot{\vec{a}} = f(t, \vec{a})$, the RK4 method
estimates $\vec{a}(t + \mathrm{d}t) \approx 
\vec{a}(t) + \mathrm{d}t \vec{b}_\text{RK4}$ with 
$\vec{b}_\text{RK4} \equiv \frac{1}{6}(\vec{b}_1 + 2\vec{b}_2 + 2\vec{b}_3 + \vec{b}_4)$
and
\begin{align*}
  \vec{b}_1 &= f(t, \vec{a}(t)), \\
  \vec{b}_2 &= f(t + \frac{\mathrm{d}t}{2}, \vec{a}(t) + \frac{\mathrm{d}t}{2} \vec{b}_1), \\
  \vec{b}_3 &= f(t + \frac{\mathrm{d}t}{2}, \vec{a}(t) + \frac{\mathrm{d}t}{2} \vec{b}_2), \\
  \vec{b}_4 &= f(t + \mathrm{d}t, \vec{a}(t) + \mathrm{d}t \vec{b}_3).
\end{align*}
The sMPS TDVP flow equations derived in the main part of this work provide
the derivative function for the $n$th site $B_n = f_n(t, [A])$, allowing us 
to implement the RK4 integrator without any additional tools. It is worth 
noting that $B_\text{RK4}$, obtained by adding the tangent vector parameters 
from the various sub-steps, is not gauge-fixing. This is because each individual 
$B_{n,i}$, although it is gauge-fixing for the sub-step point $A'$ at which it was
obtained, is not generally gauge-fixing when applied at the original point $A$. 
Additionally, each sub-step changes the gauge-choice slightly, since gauge-fixing
only holds to first order in the step size. On the other hand, since the 
gauge-fixing flow equations do preserve the gauge choice when integrated exactly,
gauge-fixing should improve with the accuracy of the numerical integration.
We should thus expect the RK4 method to maintain the gauge choice up to errors of
$\mathcal{O}(\mathrm{d}t^5)$ with each step. This is far better than the Euler
method, which incurs $\mathcal{O}(\mathrm{d}t^2)$ errors.

The error can be quantified by the change in the energy expectation 
value, which is conserved under exact time evolution. We confirm the benefits of
our RK4 implementation by comparing it to the Euler method for the Heisenberg model
example described in the main text, which we simulate on a finite chain with open 
boundary conditions in order to avoid errors due to the interface with the bulk.
To compare the efficiency of the two methods,
we set the step sizes such that the computation time per unit simulated time is
roughly the same and examine the overall change in the energy expectation value after
a period of simulated time $T$. Since a single RK4 step requires roughly four times
as much computation as an Euler step, we choose $\mathrm{d}t_\text{RK4} = 4 \mathrm{d}t_\text{Euler}$.
For $\mathrm{d}t_\text{RK4} = 0.01$, the energy errors after a time $T = 10 \hbar s$ are 
$\epsilon_\text{Euler} = -1.01 \times 10^{-3}$ and $\epsilon_\text{RK4} = -9 \times 10^{-6}$,
showing a significant advantage for the RK4 method for the same computation time.
The vast majority of the RK4 error comes from the first four steps, whereas the Euler errors
are uniformly distributed in time.
Excluding these steps from the RK4 error estimate results in $\epsilon'_\text{RK4} = -3 \times 10^{-9}$. 
Both $\epsilon'_\text{RK4}$ and $\epsilon_\text{Euler}$
are in line with the theoretical global error estimates of $\mathcal{O}(\mathrm{d}t_\text{RK4}^4)$
and $\mathcal{O}(\mathrm{d}t_\text{Euler})$ respectively. The comparatively large errors made
by the RK4 method during the first few steps are caused by the presence of
particularly small Schmidt coefficients, indicating that the bond-dimension is higher
than necessary. Small Schmidt coefficients lead to instability because the squares
of the Schmidt coefficients appear in the $l$ and $r$ matrices, which are
inverted in the TDVP algorithm, amplifying errors on small values greatly.
To mitigate this, the bond-dimension can be reduced dynamically (and increased
later if necessary), cutting off Schmidt coefficients that are close to zero.
Alternatively, an integrator that is robust under low-rank conditions could
be used \cite{lubich_projector-splitting_2013}.

\end{document}